\documentclass[twocolumn,PRL,showpacs,amsmath,amssymb,nofootinbib,aps,floatfix,notitlepage,preprintnumbers,useAMS,superscriptaddress,letterpaper,nolongbibliography]{revtex4-2}

\usepackage{aas_macros}
\usepackage{graphicx}% Include figure files
\usepackage{dcolumn}% Align table columns on decimal point
\usepackage{bm}% bold math
\newcommand{\p}{\mathcal{P}}
\newcommand{\be}{\begin{equation}}
\newcommand{\ee}{\end{equation}}
\newcommand{\fp}{\widetilde{\p}}
\newcommand{\ft}{\widetilde{T}}
\newcommand{\fpt}{\fp(\ft)}

\begin{document}

%\preprint{APS/123-QED}

\title{Breaking the intensity-bias degeneracy in line intensity mapping}

\author{Patrick C. Breysse}
\affiliation{Center for Cosmology and Particle Physics, Department of Physics, New York University, 726 Broadway, New York, NY, 10003, USA.}

\date{\today}% It is always \today, today,

\begin{abstract}
Line intensity mapping is rapidly gaining prominence as a tool to understand galaxy evolution and cosmology at high redshift.  However, the standard power spectrum analysis for these surveys suffers from a tight degeneracy between the overall mean line intensity and the bias of the emitting galaxies.  We outline a new formalism for the Voxel Intensity Distribution (VID), a one-point statistic which is known to contain information beyond the power spectrum.  We use this new calculation to show for the first time that the VID can be used to break this key degeneracy in intensity mapping data.

\end{abstract}

\maketitle

Numerous new line intensity mapping (LIM) experiments \cite{Crites2014,Keating2016,Anderson2018,Yang2019,Vieira2020,Keating2020,Crichton2021,CCAT2021,Cataldo2021,Cleary2022,CHIME2022,Karkare2022,HERA2022,Wolz2022,Anderson2022,Fasano2022} are poised to deliver a wealth of new data on the nature and evolution of high-redshift galaxies.  Rather than attempting to observe objects individually, LIM surveys map the unresolved intensity in a target spectral line over cosmological scales \cite{Kovetz2017,Kovetz2019}.  In doing so, they probe the total aggregate line emission at a given redshift, giving them insight into faint galaxy populations inaccessible to traditional methods \cite{Sun2022,Breysse2022sam,Bernal2022}.  

Depending on the target line, intensity maps contain information about many aspects of cosmology and galaxy formation.  Maps of the 21 cm fine-structure line from neutral hydrogen probe the evolving neutral gas content of the universe \cite{Wolz2022}, particularly during the Epoch of Reionization \cite{Pritchard2012}.  Other lines, like the CII fine structure line \cite{Gong2012,Silva2015,Yue2015,Dumitru2019,Padmanabhan2019,Chung2020} or rotational transitions of carbon monoxide (CO) \cite{Lidz2011,Pullen2013,Breysse2014,Li2016,Padmanabhan2018} are sourced within denser phases of the interstellar medium and are highly correlated with star formation activity.  Emitting sources trace the underlying large-scale clustering, allowing LIM data to access fundamental growth of structure physics as well \cite{Bernal2019,Dizgah2019,Liu2021,Dizgah2022a,Karkare2022a}.

As with most cosmological observables, intensity maps are most often analyzed in terms of their power spectra, which give the total power present in different Fourier modes of the line fluctuations.  The power spectrum $P(k)$ at wavenumber $k$ of most lines is typically expressed as
\be
P(k) =\overline{T}^2\overline{b}^2P_m(k)+P_{\rm{shot}}.
\label{pk}
\ee
$P_m(k)$ is the power spectrum of the underlying dark matter, $\overline{T}$ is the mean intensity of the target line, $\overline{b}$ is the luminosity-weighted bias of the emitting galaxies, and $P_{\rm{shot}}$ is the scale-independent shot noise caused by the discreteness of the emitting sources \cite{Lidz2011,Breysse2019agn}.  The form of Eq. (\ref{pk}) demonstrates a key limitation of LIM power spectra: that the mean line intensity is degenerate with the clustering bias.  $\overline{T}$ is a highly desirable quantity to measure, as it maps to the total neutral gas content of the Universe for 21 cm maps, the total molecular gas content for CO, and so on. It is almost always impossible to measure $\overline{T}$ by directly averaging a map, as large-scale information is lost during data reduction and foreground cleaning \cite{Liu2020,Foss2022}, and any measurement of $\overline{T}$ from $P(k)$ is dependent on the modeling of the line bias \cite{Beane2018,Yang2019,Wolz2022,Chung2022,Anderson2022}.  Eq. (\ref{pk}) only technically gives the monopole of the LIM power spectrum. Higher-order multipoles do include some information which weakens this degeneracy, but in practice, $\overline{T}$ and $\overline{b}$ remain quite difficult to disentangle even with anisotropies \cite{Breysse2022}.

The fluctuations seen in an intensity map are the result of highly nonlinear processes within individual galaxies, and thus are quite non-Gaussian on small scales.  The power spectrum therefore does not fully describe the intensity statistics, and we can seek other avenues to access the missing information.  Rather than cumbersome higher-order estimators, we will look at the one-point statistics of an intensity field through a quantity termed the Voxel Intensity Distribution (VID).  Based on well-established probability-of-deflection or $P(D)$ techniques \cite{Scheur1957,Barcons1992,Windridge2000,Lee2009,Patanchon2009}, the VID has been shown to be highly complementary to the power spectrum for inferring the physics behind non-Gaussian intensity maps \cite{Breysse2016,Breysse2017,Ihle2019,SatoPolito2022,Libanore2022}.  However, expressions of the intensity mapping VID used to date do not explicitly include the bias, and thus cannot self-consistently model the luminosity-clustering dichotomy at the heart of this problem.  Here we present a novel theoretical calculation of the VID which accounts for the missing information.  We then use a simple example model to demonstrate that a one-point analysis can indeed solve this key problem with the power spectrum.

First, let us consider the form of the VID used in \cite{Breysse2017} and others.  Our goal is to predict the probability\footnote{To avoid confusion, we use $P$ to refer to power spectra and $\p$ to refer to probability distributions.} $\p(T)$ of observing brightness temperature $T$ in a given voxel.  Previous works \cite{Breysse2016,Breysse2017} employed a light modification of the traditional $P(D)$ method in which
\be
\p(T)=\sum_{N=0}^\infty \p_N(T)\p(N),
\label{PofD}
\ee
where $\p(N)$ is the probability of observing $N$ galaxies in a voxel and $\p_N(T)$ is the intensity PDF given $N$, computed through successive convolution
\be
\p_N(T)=\p_{N-1}(T)\circ\p_1(T).
\label{convolve}
\ee
The one-source PDF is simply proportional to the line luminosity function.  The $P(D)$ formalism typically assumes a Poisson distribution for $\p(N)$, but because we work on cosmological scales we must account for large-scale clustering.  Ref. \cite{Breysse2017} assumed a lognormal-based model for $\p(N)$, which is known to be a reasonable if simplistic approximation to the true distribution \cite{Coles1991}.

\begin{figure*}
\centering
\includegraphics[width=\textwidth]{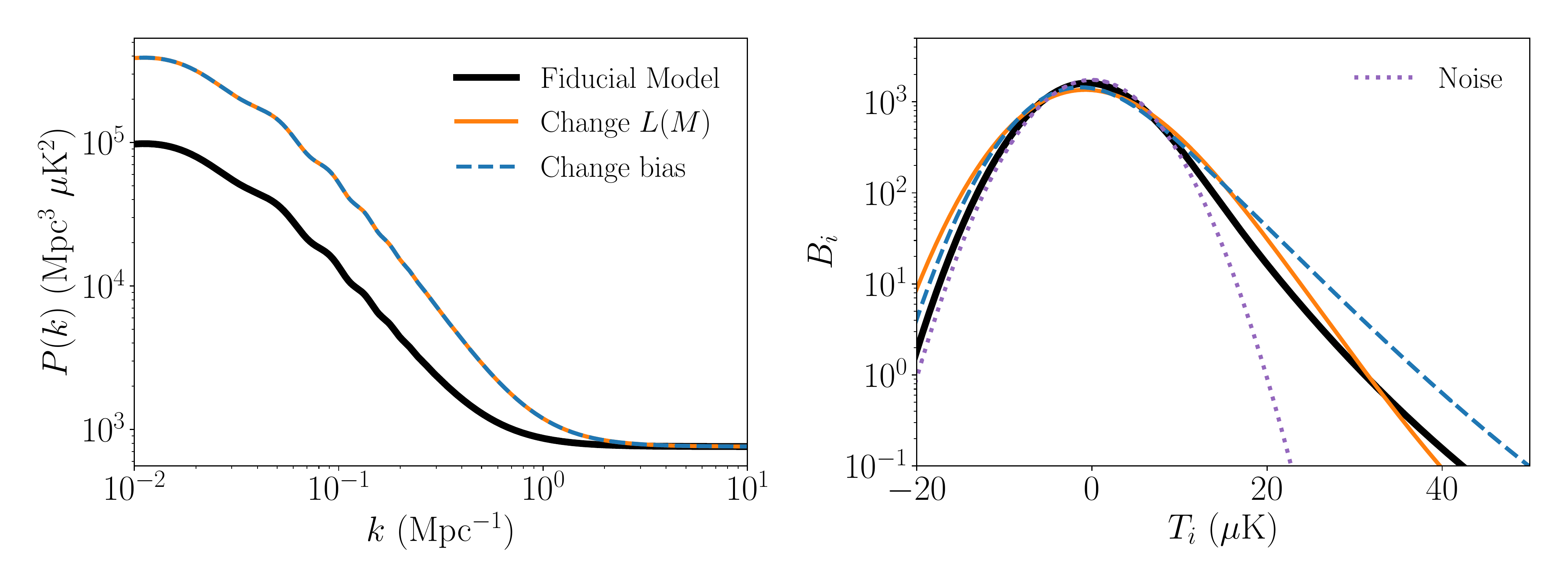}
\caption{Effect of varying $\alpha$, $\beta$, and $A_b$ on the power spectrum  (left) and VID (right, with bins of width 0.2 $\mu$K).  Thick black curves show the fiducial model, blue dashed curves double the overall galaxy bias, orange solid curves change $\alpha$ and $\beta$ in the $L_{\rm{CO}}(M)$ model to produce the same $\overline{T}\overline{b}$ and $P_{\rm{shot}}$ as the blue curve.  The contribution to the VID from instrument noise is given by the purple dotted curve.  Note that the mean intensity $\overline{T}$ has been subtracted from the histograms following Eq. (\ref{histogram}).}
\label{fig:demo}
\end{figure*}

The expressions in Eqs. (\ref{PofD}) and (\ref{convolve}) make a key assumption that renders it unsuitable for our purposes: that the luminosity of an object is independent of the density of galaxies in its vicinity.  In reality, more massive (and therefore brighter) galaxies are more biased and thus more strongly clustered.  In other words, the luminosity function in voxels with high $N$ will be skewed towards brighter objects.  In order to actually probe the relationship between clustering and the VID, we will need to take into account this crucial effect.

Our new VID calculation is based on a similar formalism applied in Refs. \cite{Thiele2019,Thiele2020} for the cosmic infrared background and weak lensing fields.  Since most lines we consider are sourced within the disks of star-forming galaxies, we model an individual line emitter as a point source residing at the center of a dark matter halo.  Satellite galaxies, which should generally be subdominant at high redshift, are left for future work.  We further assume that halos are linearly biased tracers of the underlying dark matter field. Consider first the contribution to the VID from sources with a halo mass between $M$ and $M+dM$, such that a voxel will contain exactly zero or one such objects.  The intensity PDF in a mass bin is
\be
\p(T|M)=\p(N=0)\delta_D(T)+\p(N=1)\p_1(T|M),
\ee
where $\p_1(T|M)$ is the PDF for a single source.

For reasons that will become clear, we will work with the intensity characteristic function
\be
\fp(\ft|M)=1+\p(N=1)\left(\fp_1(\ft|M)-1\right),
\ee
which is obtained by Fourier transforming the original PDF, with $\ft\equiv2\pi/T$ as the Fourier conjugate of $T$.  We have also used $\p(N=0)=1-\p(N=1)$.  Assuming Poisson statistics in our narrow mass bins, 
\be
\p(N=1)\approx\overline{N}(M)=\frac{dn}{dM}V_{\rm{vox}}\left(1+b(M)\delta_m\right)dM,
\ee
where $\overline{N}(M)\ll1$ is the mean number of halos expected in a voxel with comoving volume $V_{\rm{vox}}$, $dn/dM$ is the halo mass function, $b(M)$ is the halo bias, and $\delta_m$ is the matter density contrast at the voxel location.  Note the explicit dependence on $b(M)$, which was absent from the older VID derivation.  We can then write
\be
\fp(\ft|M)=\exp\left[p(\ft,M)\left(1+b(M)\delta_m\right)dM\right],
\ee
where for compactness we have defined
\be
p(\ft,M)\equiv\frac{dn}{dM}V_{\rm{vox}}\left(\fp_1(\ft|M)-1\right),
\ee
and used the Taylor expansion of the exponential.  Higher-order terms of order $(dM)^2$ are neglected.

The observed voxel intensity will be the sum of the contributions from all of the individual mass bins, so the total PDF will be the convolution of those of each mass bin.  By the Fourier convolution theorem, this becomes the product of the characteristic functions such that
\begin{align}
\fpt&=\prod_M \fp(\ft|M)\notag\\
&=\exp\left[\sum_Mp(\ft,M)\left(1+b(M)\delta_m\right)dM\right]\label{vid}\\
&\approx\exp\left[\int p(\ft,M)\left(1+b(M)\delta_m\right)dM\right],\notag
\end{align}
where we take $dM$ to zero to approximate the integral.

The result of Eq. (\ref{vid}) can be separated into two components.  The first,
\be
\fp_{\rm{un}}(\ft)=\exp\left[\int p(\ft,M)dM\right],
\ee
gives the distribution in the absence of clustering.  The second component encodes the clustering effect
\be
\fp_{\rm{cl}}(\ft)=\exp\left[\delta_m\int p(\ft,M)b(M)dM\right].
\ee
For our final PDF, we need to average $\fp_{\rm{cl}}$ over all realizations of $\delta_m$.  To do so, we use the identity that for Gaussian fields $\langle\exp(x)\rangle=\exp(\langle x^2\rangle/2)$ \cite{Thiele2019}.  This leaves
\be
\fp_{\rm{cl}}(\ft)=\exp\left[\frac{\sigma_m^2}{2}\left(\int p(\ft,M)b(M)dM\right)^2\right],
\ee
where
\be
\sigma_m^2=\langle\delta_m^2\rangle=\int P_m(\mathbf{k})\widetilde{W}_{\rm{vox}}^2(\mathbf{k})\frac{d^3\mathbf{k}}{(2\pi)^3},
\ee
is the variance of the matter contrast on the scale of a voxel with window function $W_{\rm{vox}}$.  Finally, we can add in the effect of instrument noise by convolving in the noise PDF $\p_{\rm{Noise}}(T)$.  The final observed VID can then be obtained by inverse Fourier transforming
\be
\fp_{\rm{obs}}(\ft)=\fp_{\rm{un}}(\ft)\fp_{\rm{cl}}(\ft)\fp_{\rm{Noise}}(\ft).
\ee

We demonstrate our formalism with a simple example forecast based on the CO Mapping Array Project (COMAP, \cite{Cleary2022}).  Note that our goal is not to perform a rigorous experimental forecast, but to explore the differences between bias and luminosity in the VID.

Specifically, we imagine an intensity map targeting the 115 GHz CO(1-0) line using the second-generation COMAP-EoR survey with instrument specifications given in Ref. \cite{Breysse2022}.  COMAP-EoR will map the CO(1-0) line in redshift bands between $z=2.6$ and 8.6, along with some CO(2-1) emission that we neglect here.  To model the CO luminosity we apply the scaling relation method from Refs. \cite{Li2016,Keating2020}, referred to in Ref. \cite{Breysse2022} as the ``Li16/Keating20" model.  In this model, we relate the CO luminosity $L_{\rm{CO}}$ of a halo to its infrared luminosity $L_{\rm{IR}}$ by
\be
\log L'_{\rm{CO}}(M)=\alpha\log L_{\rm{IR}}(M)+\beta,
\label{LofM}
\ee
where $x'\equiv L_{\rm{CO}}/ L'_{\rm{CO}}=4.9\times10^{-5}\ L_{\odot}/{(\rm K\ km\ s^{-1}\ pc^2)}$ is the conversion between physical and observer units \cite{Li2016}.  We use the fits from Ref. \cite{Kamenetzky2016} for $\alpha$ and $\beta$.  Ref. \cite{Li2016} provides a model for $L_{\rm{IR}}(M)$ \cite{Behroozi2013}.  We make the common assumption of a lognormal scatter around the mean relation from Eq. (\ref{LofM}) with width $\sigma_{\rm{sc}}=0.37$ dex.

Under this model, we can write the components of the power spectrum (Eq. \ref{pk}) as
\be
\overline{T}=C_{\rm{LT}}\int_{M_{\rm{min}}}^\infty L_{\rm{CO}}(M)\frac{dn}{dM}dM,
\ee
\be
\overline{b}=\frac{\int_{M_{\rm{min}}}^\infty L_{\rm{CO}}(M)b(M)dn/dM dM}{\int_{M_{\rm{min}}}^\infty L_{\rm{CO}}(M)dn/dM dM},
\ee
and
\be
P_{\rm{shot}}=C_{\rm{LT}}^2e^{(\sigma_{\rm{sc}}\ln10)^2}\int_{M_{\rm{min}}}^\infty L_{\rm{CO}}^2(M)\frac{dn}{dM}dM,
\ee
where $M_{\rm{min}}=10^{10}\ M_{\odot}$ is the minimum mass for an emitting galaxy. $C_{\rm{LT}}=c^3(1+z)/8\pi k_B\nu^3H(z)$ converts between luminosity density and brightness temperature for a line with rest frequency $\nu$ at redshift $z$, where $c$ is the speed of light, $k_B$ is Boltzmann's constant, and $H(z)$ is the Hubble parameter.  We can also write the full lognormal expression for the one-source intensity PDF
\be
\p_1(T|M)=\frac{1}{T\sigma_{\rm{sc}}\sqrt{2\pi}\ln10}e^{-(\log(T/X_{LT})-\mu)^2/2\sigma_{\rm{sc}}^2},
\ee
where $X_{LT}=C_{LT}/V_{\rm{vox}}$ converts luminosity to brightness temperature in a voxel with volume $V_{\rm{vox}}$ \cite{Breysse2017} and
\be
\mu\equiv\log\left(L_{\rm{CO}}(M)\right)-\frac{1}{2}\sigma_{\rm{sc}}^2\ln(10),
\ee
sets the mean of the distribution to $L_{\rm{CO}}(M)$.

We parameterize our lack of knowledge of the bias by
\be
b(M)=A_bb_0(M),
\label{bofM}
\ee
where we take the Tinker form of $b_0(M)$ \cite{Tinker2010} and the corresponding mass function $dn/dM$ \cite{Tinker2008}. The actual VID estimator we will work with will be the map histogram
\be
B_i=N_{\rm{vox}}\int_{T_i} \p(T-\overline{T})dT,
\label{histogram}
\ee
which gives the expected number of voxels with intensity in a bin given by $T_i$.  As in Ref. \cite{Breysse2017}, we assume the overall mean of the map has been subtracted.  This will approximately account for the loss of large-scale information due to foregrounds and systematics \cite{Ihle2022}.  We model noise on the power spectrum monopole following Ref. \cite{Breysse2022}, and assume a Gaussian white noise contribution to the VID based on the same instrument model.

Figure \ref{fig:demo} shows the effects of varying $\alpha$, $\beta$, and $A_b$ on the power spectrum and VID.  We plot three models: our fiducial model, one where we double $A_b$, and another where we choose $\alpha$ and $\beta$ to produce the same $\overline{T}\overline{b}$ and $P_{\rm{shot}}$ as the model with $A_b=2$.  We can see that the latter two power spectra are identical, but the VIDs differ substantially, showing that the VID does contain additional information.

\begin{figure}
\centering
\includegraphics[width=\columnwidth]{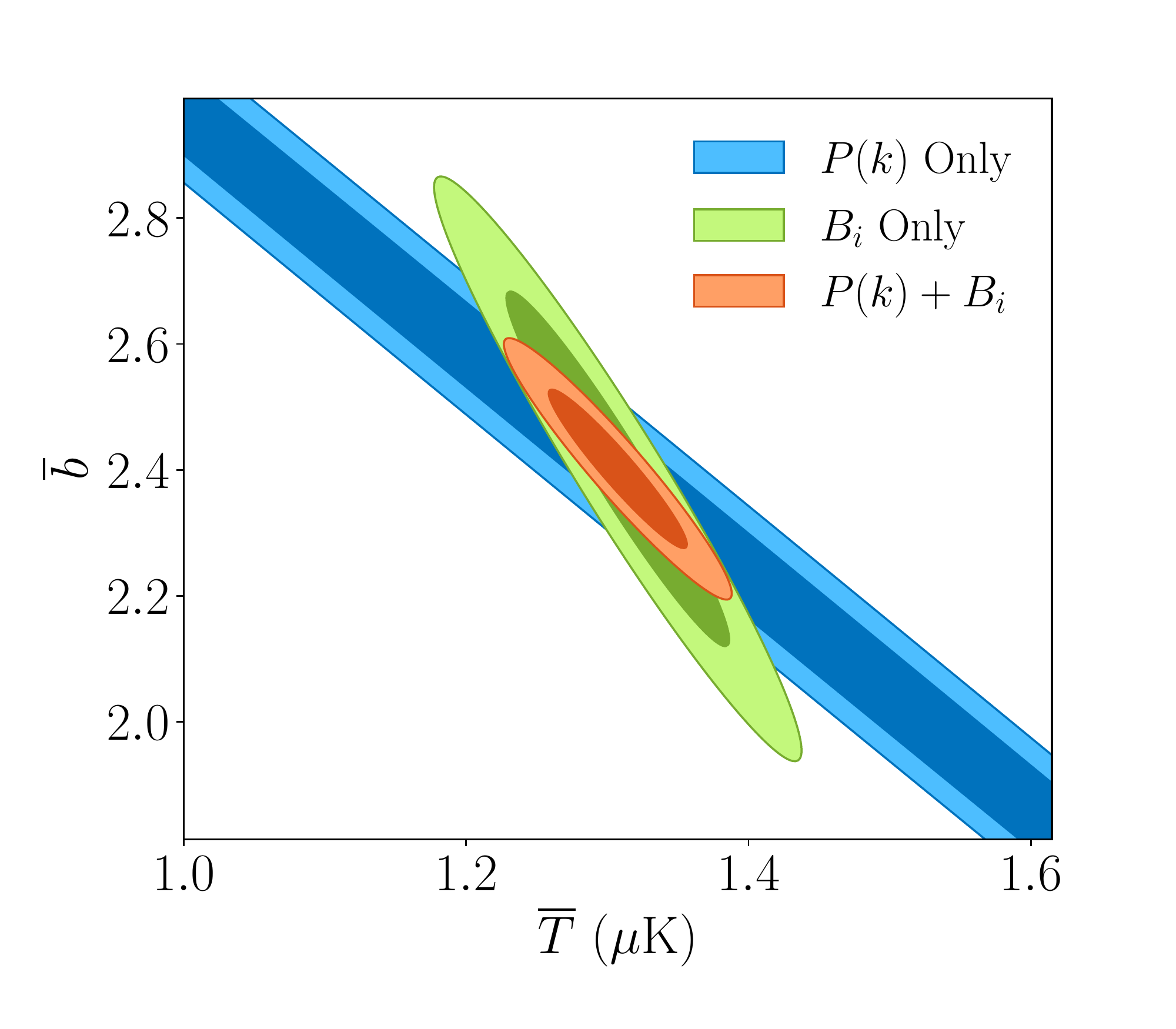}
\caption{Results of our Fisher forecast for the intensity-bias degeneracy.  We compute constraints on $\alpha$, $\beta$, and $A_b$ from the power spectrum (blue), VID (green), and their combination (orange), convert them to constraints on $\overline{T}$, $\overline{b}$, and $P_{\rm{shot}}$ while marginalizing the latter.  Constraints assume flat, uninformative priors on all quantities and use the $26-30$ GHz observed frequency range from COMAP-EoR \cite{Breysse2022}, corresponding to $z=2.4-2.8$ for CO(1-0).}
\label{fig:fisher}
\end{figure}

\begin{figure}
\centering
\includegraphics[width=\columnwidth]{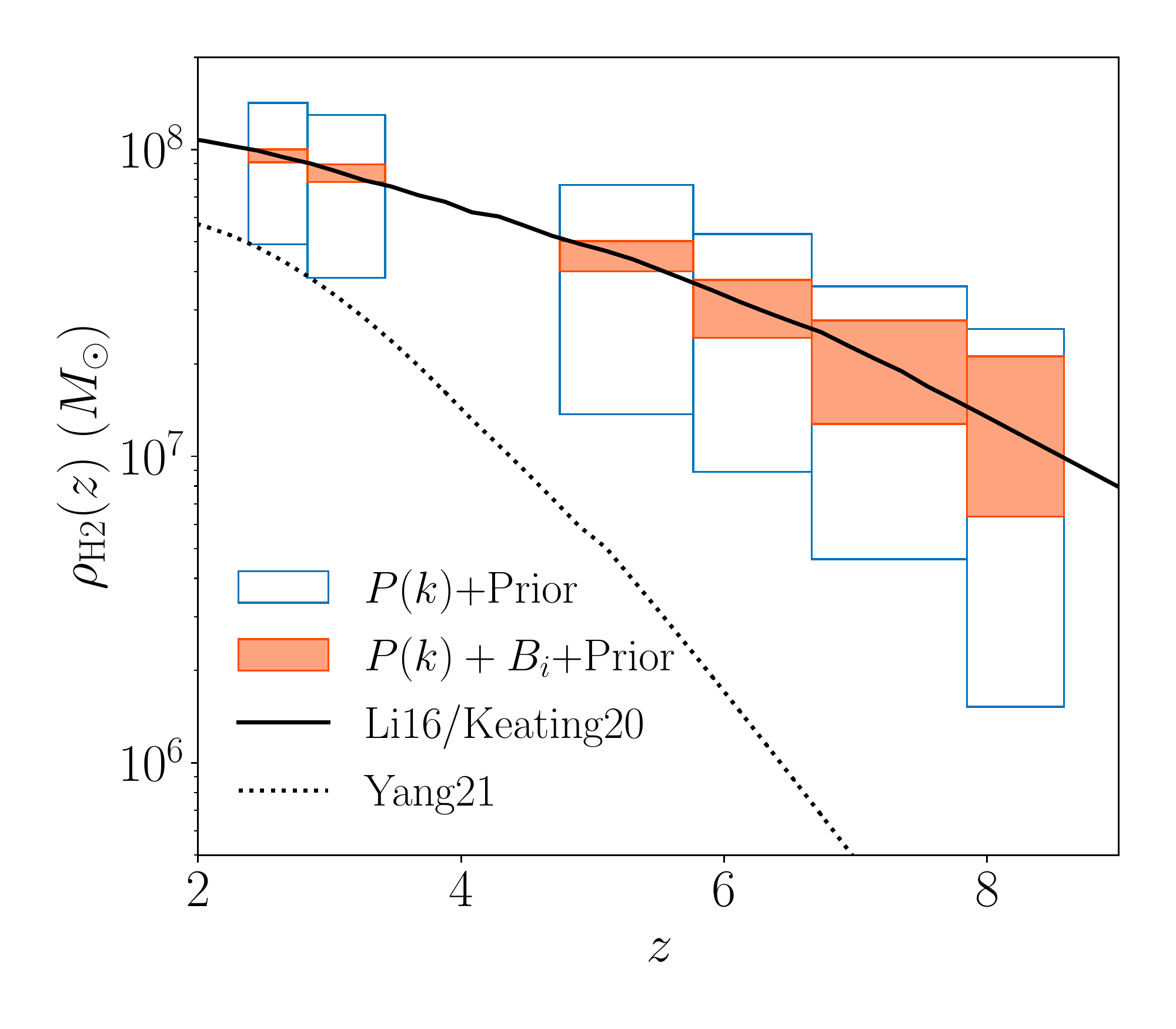}
\caption{Constraints on the cosmic molecular gas history assuming our example model.  Black lines show the gas density under our fiducial Li16/Keating20 model (solid) and for comparison the fainter Yang22 model (dotted).  Blue empty boxes show power spectrum-only constraints using the bias prior estimated from the standard deviation of the $\overline{b}$ values of the models presented in Ref. \cite{Breysse2022}. Orange filled boxes show the improvement from adding the new VID formalism.}
\label{fig:rhoh2}
\end{figure}

We carry out a Fisher matrix forecast over $\alpha$, $\beta$, and $A_b$ for our two observables and their combination.  We assume binomial errors on $B_i$ following Refs. \cite{Ihle2019,SatoPolito2022}, and we model the error on the power spectrum monopole following Refs. \cite{Bernal2019,Breysse2022}. We then convert the constraints on these parameters to constraints on $\overline{T}$, $\overline{b}$, and $P_{\rm{shot}}$ and marginalize over the shot noise.  We start by assuming uninformative flat priors on all parameters.  The result is shown in Figure \ref{fig:fisher} for the lowest redshift band of COMAP-EoR, with observed frequencies 26-30 GHz corresponding to $z=2.4-2.8$ for CO(1-0).  We can clearly see the strong (in this approximation, total) degeneracy between $\overline{T}$ and $\overline{b}$ from $P(k)$ alone, and how the VID allows the quantities to be measured separately.

It may seem odd that we can measure $\overline{T}$ at all from the VID when we have subtracted the map mean from Eq. (\ref{histogram}).  This is because we fit for $\alpha$ and $\beta$ and converted the results to $\overline{T}$.  Neither $\alpha$ nor $\beta$ are degenerate with a constant subtraction, so we get a useful constraint.  This effect is discussed in more detail in Ref. \cite{Breysse2017}.

Note that we treat $P(k)$ and $B_i$ as independent, though they are known to be mildly correlated \cite{Ihle2019}.  A derivation of this covariance for the older VID formalism exists \cite{SatoPolito2022}, we leave an update to that calculation for future work.  The covariance will weaken the joint constraints in Fig. \ref{fig:fisher} somewhat, but it should not change the basic fact of the broken degeneracy.

To illustrate the importance of this result, we plot constraints on the cosmic molecular gas history using our simple model.  For this example, we would like a more useful prior on the bias amplitude.  Ref. \cite{Breysse2022} examines seven CO emission models.  We assume a (very rough) Gaussian prior on $A_b$ with width set by the standard deviation of the $\overline{b}$ values of these models.  We compute constraints on $\rho_{\rm{H2}}$ using
\be
\rho_{\rm{H2}}=\frac{\alpha_{\rm{CO}}}{x'}\int L_{\rm{CO}}(M)\frac{dn}{dM}dM
=\frac{\alpha_{\rm{CO}}}{x'C_{LT}}\overline{T},
\ee
where $\alpha_{\rm{CO}}$ is the ratio between a galaxy's molecular gas mass and its CO(1-0) luminosity.  We take the typical high-redshift value of $\alpha_{\rm{CO}}=3.6\ M_{\odot}/($K km/s pc$^2)$ \cite{Daddi2010}. Figure \ref{fig:rhoh2} shows the result of this forecast for the six redshift bands where COMAP-EoR accesses the CO(1-0) line.  It can clearly be seen that, by breaking the $\overline{T}$--$\overline{b}$ degeneracy, the VID has significantly improved the constraining power under this model.

Our demonstration here, while it illustrates a clear new utility of the VID, is by no means a complete exploration of the relevant model dependencies.  The two-parameter $L_{\rm{CO}}(M)$ model is likely too simple to capture important realities of the actual signal \cite{Padmanabhan2018,Chung2022}, and our simple treatment of the bias in Eq. (\ref{bofM}) belies the fact that $b(M)$ in practice depends on the broader cosmological model in ways which are not limited to an overall normalization.  We have neglected important effects like the aforementioned covariance between $P(k)$ and $B_i$, but also the presence of satellite galaxies \cite{Schaan2021}, redshift-space distortions \cite{Kaiser1987,Bernal2019}, and other higher-order clustering effects \cite{Dizgah2022}.  We have also ignored other contaminating emission, including continuum foregrounds \cite{Foss2022} and the CO(2-1) line \cite{Breysse2022}.  The importance of these approximations will need to be assessed in future work.

Despite the simplicity of our model though, it is quite clear that the information contained in the VID accomplishes our goal of separating the clustering and luminosity information in the power spectrum.  For LIM measurements which mix together information from parsec to gigaparsec scales, having the ability to isolate specific aspects of the emitting population will be key.  Our results here serve to firmly demonstrate the value of the VID analysis to understand the high-redshift universe from the rapidly-growing volume of intensity mapping data.

This work was supported by the James Arthur Postdoctoral Fellowship.  The author would like to thank Dongwoo Chung, Anthony Pullen, H\aa vard Ihle, Delaney Dunne, and Kieran Cleary for useful conversations and comments on the manuscript.

\bibliography{references}% Produces the bibliography via BibTeX.

\end{document}